\pdfoutput=1

\documentclass{llncs}
\usepackage{authblk}
\usepackage{tikz}
\usetikzlibrary{trees}

\usepackage{amsthm}

\usepackage[linesnumbered,ruled]{algorithm2e}
\usepackage{float}
\floatstyle{plaintop}
\restylefloat{table}
\usepackage[font=small]{caption}
\usepackage{amssymb}

\RequirePackage{filecontents}

\usepackage{graphicx}

\usepackage[section]{placeins}
\usepackage[utf8]{inputenc}
\usepackage[english]{babel}
\usepackage{booktabs,multirow,widetable}
\usepackage{mathtools}
\usepackage{lastpage}

\usepackage{fancyhdr}
\usepackage[page]{totalcount}

\usepackage{float}
\restylefloat{table}
\usepackage{cite}

\usepackage{filecontents}
\usepackage{hyperref}
\usepackage{amssymb}
\usepackage{etoolbox}
\AtEndEnvironment{quote}{\centering}

\theoremstyle{definition}

\newtheorem{fact}{Fact}

\usepackage{lipsum}

\usepackage{mathtools}
\DeclarePairedDelimiter{\ceil}{\lceil}{\rceil}
\usepackage{framed}
\definecolor{shadecolor}{RGB}{190,190,190}

\theoremstyle{definition}
\newtheorem*{theorem*}{Theorem}

\makeatletter
\let\scshape\relax % to avoid a warning
\DeclareRobustCommand\scshape{%
  \not@math@alphabet\scshape\relax
  \ifnum\pdf@strcmp{\f@family}{\familydefault}=\z@
    \fontfamily{qbk}%
  \fi
  \fontshape\scdefault\selectfont}
\makeatother

\newcommand{\block}[2]{%
  \begin{array}[t]{@{}c@{}}
  \boxed{\mathstrut #1}\\
  \scriptstyle\blockleftarrowfill\,#2\,\blockrightarrowfill
  \end{array}%
}
\newcommand{\blockleftarrowfill}{%
 \mathord\leftarrow
 \mkern -7mu
 \cleaders\hbox{$\scriptstyle\mkern -2mu\smash-\mkern -2mu$}\hfill
 \mkern -7mu\smash-%
}
\newcommand{\blockrightarrowfill}{%
  \smash-%
  \mkern -7mu
  \cleaders\hbox{$\scriptstyle\mkern -2mu\smash-\mkern -2mu$}\hfill
  \mkern -7mu
  \mathord\rightarrow
}

\title{Space Efficient Representations of Finite Groups}

\author{Bireswar Das$^1$  Shivdutt Sharma$^1$ \ and P.R.Vaidyanathan$^2$ \thanks{Part of this work was done when the third author was an M.Tech student at IIT Gandhinagar.} }
 \institute{$^1$IIT Gandhinagar, India \\ \{bireswar, shiv.sharma\}@iitgn.ac.in\\ $^2$TU Wien, Vienna, Austria\\vaidyanathan@ac.tuwien.ac.at}
 
\date{
    $^1$IIT Gandhinagar, India\\
    $^2$TU Wien, Vienna, Austria\\[2ex]%
    \today
}

\begin{document}

\maketitle

\begin{abstract}
The Cayley table representation of a group uses $\mathcal{O}(n^2)$ words for a group of order $n$ and answers multiplication queries in time $\mathcal{O}(1)$. It is interesting to ask if there is a $o(n^2)$ space representation of groups that still has $\mathcal{O}(1)$ query-time. We show that for any $\delta$, $\frac{1}{\log n} \le \delta \le 1$, there is an $\mathcal{O}(\frac{n^{1 +\delta}}{\delta})$ space representation for groups of order $n$ with $\mathcal{O}(\frac{1}{\delta})$ query-time.\medskip

We also show that for Z-groups, simple groups and several group classes defined in terms of semidirect product, there are linear space representations with at most logarithmic query-time.\medskip

Farzan and Munro (ISSAC'06) defined a model for group representation and gave a succinct data structure for abelian groups with constant query-time. They asked if their result can be extended to categorically larger group classes. We construct data structures in their model for   Hamiltonian groups and some other classes of groups with constant query-time.  

\let\thefootnote\relax\footnotetext{\begin{itemize}
    \item A preliminary version of this article appeared in the 22nd International Symposium on Fundamentals of Computation Theory, FCT 2019.
    \item This project is partially supported by ACM India-IARCS.
\end{itemize}}

\end{abstract}

\section{Introduction}

Groups are important algebraic structures which can be used to study symmetries in objects. Group theory has many important applications in physics, chemistry, and materials science. Group theory has been used elegantly in proving various important results in computer science, such as Barrington's theorem \cite{barrington1989bounded}, results on the graph isomorphism problem \cite{babai2016graph,luks1982isomorphism} etc. 

Algorithms for computational group theoretic problems are essential building blocks for many of the computer algebra systems such as GAP, SageMath, Singular etc. Some of the fundamental group theoretic algorithms were designed based on the ideas of Sims and Schreier (see \cite{seress2003permutation}). Various computational group theoretic problems such as the group isomorphism problem, set stabilizer problem for permutation groups are also interesting from a purely complexity theoretic point of view for their connection with the graph isomorphism problem \cite{eugene1993permutation}.

Two of the most commonly used ways of representing groups are via generators in the permutation group setting and via Cayley tables. Several interesting problems such as the group isomorphism problem, various property testing problems, the group factoring problem etc., have been studied for groups represented by their Cayley tables \cite{Kav,Kay,Vik,che,Sar,Wil,kumar1998property}.

While a multiplication query for a group of order $n$  can be answered in constant time in the Cayley table representation, the space required to store the table is $\mathcal{O}(n^2 \log n)$ bits or $\mathcal{O}(n^2)$ words in word-RAM model, which is prohibitively large. It is interesting to know if there are data structures to store a group using $o(n^2)$ words but still supporting constant query-time for multiplication. We construct a data structure that has constant query-time but uses just $\mathcal{O}(n^{1.05})$ words to represent the group. In fact, our result is more general and offers several other interesting space versus query-time trade-offs.

 We note that there are space efficient representations of groups such as the generator-relator representation (see \cite{magnus2004combinatorial}), polycyclic representation \cite{sims1994computation} etc., that store groups succinctly. However, answering multiplication queries generally takes too much time. For example with a polycyclic representation of a cyclic group it takes linear time to answer a multiplication query. 
 
 An easy information theoretic lower bound \cite{farzan2006succinct} states that to represent a group of order $n$, at least $n\log n$ bits (or $\Omega(n)$ words in word RAM model) are needed. We do not know if in general it is possible to use only $\mathcal{O}(n)$ words to store a group while supporting fast query-time. We show that for restricted classes of groups such as Z-groups, simple groups it is possible to construct data structures that use only $\mathcal{O}(n)$ space and answer multiplication query in $\mathcal{O}(1)$ and $\mathcal{O}(\log n)$ time respectively.

  In the past space succinct representation of groups has been studied for restricted classes of groups 
  \cite{farzan2006succinct,leedham1990collection}. Farzan and Munro \cite{farzan2006succinct} defined a model of computation in which a compression algorithm is applied to the group to get a succinct canonical form for the group. The query processing unit in their model assumes that the group elements to be multiplied are in given by their labels. They also assume that the query processing architecture supports an extra operation called bit-reversal. In their model they show that for abelian groups, the query processing unit needs to store only constant number of words in order to answer multiplication queries in constant time. Farzan and Munro ask if their results can be extended to  categorically larger classes of groups. We show that we can design space efficient data structures with same space bounds and query-time for Hamiltonian groups and Z-groups.
  Hamiltonian groups are nonabelian groups all of whose subgroups are normal. Z-groups are groups all of whose Sylow subgroups are cyclic. There are many interesting nonabelian groups in the class of Z-groups.   
  We also show that in their model constant query-time can be achieved for larger classes of groups defined in terms of semidirect products provided the query processing unit is allowed to use linear space.

%##########################################################################

   %                           PRELIMINARY SECTION

%##########################################################################

\section{Preliminary}\label{preliminarary}

 In this section, we describe some of the group-theoretic definitions and background used in the paper. For more details see  \cite{Hal,che,dummit2004abstract,erdos1965probabilistic}.
 
 For a group $G$,  the number of elements in $G$ or the \emph{order} of $G$ is denoted by $\vert G \vert$.  Let $x \in G$ be an element of group $G$, then ord$_{G}(x)$ denotes the order of the element $x$ in $G$, which is the smallest power  $i$ of $x$ such that $x^i =e$, where $e$ is the identity element of the group $G$.
  For a subset $S \subseteq G$, $\langle S \rangle$ denotes the subgroup generated by the set $S$. 
  
  A group \emph{homomorphism} from $(G,\cdot)$ to $(H, \times)$ is a function $\varphi :  G \longrightarrow H$ such that $\forall g_1,g_2 \in G, \varphi(g_1 \cdot g_2) = \varphi(g_1) \times \varphi(g_2)$. A bijective homomorphism is called an \emph{isomorphism}. Let $\text{Aut}(H)$ denote  \emph{automorphism} group of $H$, Aut$(H) = \{\sigma \mid \sigma : A \longrightarrow A \text{ is an isomorphism } \}$. The set of all automorphism from a $H$ to $H$ under function composition forms a group. Two elements $a$ and $b$ with the conditions, $a^4 =1,a^2=(ab)^2=b^2$ generates a nonabelian group of order $8$ known as $\emph{quaternian group}$. A group is said be a \emph{simple} if every non-trivial subgroup of it is not a normal subgroup.  Let $G$ be a finite group and $A ,B$ be subgroups of $G$. Then $G $ is a \emph{direct product} of $A$ and $B$, denoted $G= A \times B$, if 1) $A\mathrel{\unlhd}G$ and $B\mathrel{\unlhd}G$, 2) $\vert G \vert= \vert A \vert \vert B \vert$, 3) $A \cap B =\{e\}$.

  Let $A$ and $B$ be two groups and let $\varphi:B\longrightarrow \text{Aut}(A)$ be a homomorphism. The semidirect product of $A$ and $B$ with respect to $\varphi$, denoted $A \rtimes_{\varphi} B$, is a group whose underlying set is $A\times B$ and the group multiplication is define as follows: Let $(a_1,b_1),(a_2,b_2)\in A\times B$. The multiplication of $(a_1,b_1)$ and $(a_2,b_2)$  is defined as $(a_1(\varphi(b_1)(a_2)),b_1b_2)$. It is routine to check that the resulting structure is indeed a group. A group $G$ is called the semidirect product of two of its subgroups $A$ and $B$ if there exists $\varphi:B\longrightarrow \text{Aut}(A)$ such that $G\cong A \rtimes_{\varphi} B$.

  A group $G$ is said to be \emph{abelian} if $ab =ba, \forall a,b \in G$. The fundamental theorem for finitely generated abelian groups implies that a finite abelian group $G$ can be decomposed as a direct product $G=G_1 \times G_2 \times \ldots \times G_t$, where each $G_i$ is a cyclic group of order $p^j$ for some prime $p$ and integer $j \ge 1$. If $a_i$ generates the cyclic group $G_i$ for $i =1,2,3,\ldots,t$ then the elements $a_1 , a_2 , \ldots , a_t$ are called a \textit {basis} of G.

A group $H$ is \emph{Hamiltonian} if every subgroup  of $H$ is normal. It is a well known fact that \cite{Hal} a group is Hamiltonian if and only if  \\
- $G$ is the quaternion group $Q_8$; or,\\
- $G$ is the direct product of $Q_8$ and $B$, of $Q_8$ and $A$, or of $Q_8$ and $B$ and $A$,\\
where $A$ is an abelian group of odd order \textbf{$k$} and $B$ is an elementary abelian $2$-group.\footnote{An elementary abelian $2$-group is an abelian group in which every nontrivial element has order $2$.}  
A group is \emph{Dedekind} if it is either abelian or Hamiltonian.

 Let $p^k$ is the highest power of a prime $p$ dividing the order of a finite group $G$,  a subgroup of  $G$ of order $p^k$ is called a \emph{Sylow $p$-subgroup} of $G$. 
 
 Z-groups are groups all of whose Sylow subgroups are cyclic. A group $G$ is a Z-group if and only if it can be written as a semidirect product of two cyclic groups.

Let $G$ be a group with $n$ elements. A sequence $(g_1,\ldots,g_k)$ of $k$ group elements is said to be \emph{cube generating set}  of $G$ if 

\begin{equation}
    G= \{g_1^{\epsilon_1}g_2^{\epsilon_2}\cdots g_k^{\epsilon_k}\mid \epsilon_i \in \{0,1\}\}
\end{equation}

Let $G=\langle S\rangle$. The \emph{Cayley graph} of the group $G$ on generating set $S$ is the directed graph $X = (V,E)$ where $V=G$ and $E=\{(g,gs) \mid g\in G, s\in S\}$. Additionally, every edge $(g,gs)  ~\forall g \in G,~\forall s \in S$ is labeled with $s$. We denote $\text{diameter}(G, S)$ as the graph diameter of the Cayley graph of group $G$ on generating set $S$.

We use the $\mathbin{\%}$ symbol as the modulo operator such that $a \mathbin{\%} b$ denotes the remainder of $a$ when divided by $b$.

\subsection{Model of Computation}
 The model of computation we follow is the word RAM model, where random access can be done in constant time. Each register and memory unit can store $\mathcal{O}(\log n)$ bits where $n$ is the input size (in our case $n$ is the order of the given group). These memory units are called words. The arithmetic, logic and comparison operations on $\mathcal{O}(\log n)$ bits words take constant time.  Unless stated otherwise we assume that the elements of the group are encoded as $1, 2, \ldots , n$. 
 
 The group $G$ and its Cayley table are already known and we are allowed to preprocess the input in finite time in order to generate the required data structures for the multiplication operation.  The time and space required in the preprocessing phase is not counted towards the space complexity and query time of the data structure. 
 
 In many of the existing results the time taken to generate space efficient data structures could be, for example, exponential. However, for some of our data structures the preprocessing time is polynomial in the size of input group.

  The space complexity is measured in terms of the number of words required for the storage of the data structure. The \emph{multiplication query} for a group $G$ takes two elements $x$ and $y$, and it has to return  $z=xy$. 
 
 We note that in this model the inverses of each element can be trivially stored in $\mathcal{O}(n)$ space. Thus, we primarily focus on the problem of answering multiplication queries using data structures that uses less space.

%##########################################################################
   %                        OUR  RESULTS

%##########################################################################

\section{Our Results}

In Section 4 and 5 we present space efficient data structures for various group classes in standard word RAM model. Our results for various group classes are summarized  below. We start with a representation for general groups and then  move towards more restricted group classes such as Z-groups, simple groups etc..

\begin{theorem}\label{main theorem}
Let $G$ be a group of order $n$. Then for any $\delta$ such that $\frac{1}{\log n} \le \delta \le 1$, there is a representation of $G$ that uses $\mathcal{O}(\frac{n^{1+\delta}}{\delta})$ space and answers multiplication queries in time $\mathcal{O}(\frac{1}{\delta})$. 
\end{theorem}

\begin{theorem}\label{Z-groups}

There is a representation of Z-groups such that multiplication operation can be performed using $\mathcal{O}(n)$ space in $\mathcal{O}(1)$ time.

\end{theorem}

\begin{theorem}\label{simple-groups}
There is a representation of simple groups using $\mathcal{O}(n)$ space such that multiplication query can be answered in $\mathcal{O}(\log n)$ time.
\end{theorem}

In Section 6 we study space efficient  representation of groups in the model defined by Farzan and Munro \cite{farzan2006succinct}. Our results in this model are listed below.

\begin{theorem}\label{hamiltonian-munro}
There is a representation of Hamiltonian groups such that multiplication operation can be performed using $\mathcal{O}(1)$ space in $\mathcal{O}(1)$ time.
\end{theorem}

\begin{theorem}\label{Z-groups-munro}
There is a representation of Z-groups such that multiplication operation can be performed using $\mathcal{O}(1)$ space in $\mathcal{O}(1)$ time.
\end{theorem}

\begin{theorem}\label{semi-munro}
There is a representation of groups $G = A \rtimes C_m$ such that multiplication operation can be performed in using $\mathcal{O}(\vert A \vert)$ space in $\mathcal{O}(1)$ time.
\end{theorem}

\section{Space Efficient Representations of Finite Groups}
In this section we construct a space efficient representation of a given  finite group that can quickly answer multiplication queries. More precisely,  

{
\renewcommand{\thetheorem}{\ref{main theorem}}
\begin{theorem}
Let $G$ be a group of order $n$. Then for any $\delta$ such that $\frac{1}{\log n} \le \delta \le 1$, there is a representation of $G$ that uses $\mathcal{O}(\frac{n^{1+\delta}}{\delta})$ space and answers multiplication queries in time $\mathcal{O}(\frac{1}{\delta})$. 
\end{theorem}

}

One of the main ingredients in the proof of the theorem is the existence of short cube generating sequence. Erd\"os and Renyi  \cite{erdos1965probabilistic} showed that for any group of $G$ of order $n$, there are cube generating sets of length $\mathcal{O}(\log n)$. The next theorem states this fact more formally.

\begin{theorem}[\cite{erdos1965probabilistic}]
Let $G$ be a finite group of order $n$. Then there is a sequence $(g_1,\ldots,g_k)$ of elements of $G$ such that 
\begin{enumerate}
    \item $k = \mathcal{O}(\log n)$
    \item Every element $g$ of $G$ can be written as $g_1^{\epsilon_1}g_2^{\epsilon_2}\ldots g_k^{\epsilon_k}$ where $\epsilon_i \in \{0,1\}, \forall i \in k$
\end{enumerate}
\end{theorem}

In Subsection~\ref{Deterministic Polynomial Time Preprocessing} we give a deterministic polynomial time algorithm to compute a cube generating sequence of logarithmic length. The algorithm is based on the ideas presented by Babai and Erd\"os \cite{babai1982representation}. The algorithm also servers as a proof of the above theorem.

\begin{proof}(Proof of Theorem~\ref{main theorem}) Let $\{g_1g_2\ldots g_k\}$ be a cube generating set for the group $G$. A cube generating set could be found by brute-force. For each $g\in G$ and $i\in [k]$ we fix $\epsilon_i(g)$  such that $g=\prod_{i=1}^kg_i^{\epsilon_i(g)}$.

Let $h\in G$. To compute the product $hg$, we first compute $x_1=hg_1^{\epsilon_1(g)}$. Inductively, we compute $x_i=x_{i-1}g_i^{\epsilon_i(g)}$. Here $x_k=hg$. Note that $g_i^{\epsilon_i(g)}$ is either $g_i$ or identity. In the later case there is actually no multiplication. With suitable data structures this method has query time $\mathcal{O}(k)$. However, to obtain a general result that gives interesting space versus query-time trade-offs we take the following route.

First, divide the $k$-length sequence $g_1^{\epsilon_1(g)}g_2^{\epsilon_2(g)}\cdots g_k^{\epsilon_k(g)}$ into $l$ sized blocks as shown below.

\[
 g = \block{\ g_1^{\epsilon_1(g)} \cdots\  g_l^{\epsilon_l(g)}\ }{l}\quad
   \block{g_{l+1}^{\epsilon_{l+1}(g)} \cdots g_{2l}^{\epsilon_{2l}(g)}}{l}\quad
   \cdots\quad
   \block{g_{r}^{\epsilon_{r}(g)}\cdots g_k^{\epsilon_k(g)}\quad}{l}
\]

There are $2^l$ possible products in each block and each such product will be an element of the group $G$. We will store the result of the multiplication of every element $g \in G$ with each possible $l$-length combination from each block. Each element $g$ can be  seen as a sequence of $m$  words $w_1(g),\ldots,w_m(g)$, where $m = \left\lceil \frac{k}{l}\right\rceil$ and
\begin{equation}
    w_i(g)=\prod_{j=(i-1)\ell+1}^{i\ell}g_j^{\epsilon_j(g)}
\end{equation}

Let $s_i(g)$ be the number whose binary representations is $\epsilon_{(i-1)l+1}(g)\ldots \epsilon_{il}(g)$. The number $s_i(g)$ can be viewed as a representation  of the word $w_i(g)$.

\paragraph{Data structures:} In order to perform the multiplication, we will use the following data structures which are constructed during the preprocessing phase.
\begin{enumerate}
    \item Word Arrays: For each $g\in G$ an array $\mathcal{W}_g$ of length $m$. The $i$th element $\mathcal{W}_g[i]$ in the array is set to $s_i(g)$. 

    \item Multiplication  Arrays: For each $g\in G$ and $i\in [m]$ an array $\mathcal{A}_{g}^{(i)}$ of length $2^l$. The $j$th element of $\mathcal{A}_g^{(i)}$ is computed as follows. First we compute the binary representation $\epsilon_1\epsilon_2\ldots \epsilon_l$ of $j-1$ (possibly padding 0's in the left to make it an $l$-bit binary number). We set $\mathcal{A}_{g}^{(i)}[j] = g ~ g_{(i-1)l+1}^{\epsilon_1}g_{(i-1)l+2}^{\epsilon_2}\ldots g_{il}^{\epsilon_l}$.

\end{enumerate}

\paragraph{Query Time:} Given $h$ and $g$, we want to compute $hg$. First we obtain the sequence $s_1(g),\ldots,s_m(g)$ from the word array $\mathcal{W}_g$. By design, this sequence corresponds to $w_1(g),\ldots, w_m(g)$ and $g=w_1(g)\ldots w_m(g)$. Now access array $\mathcal{A}_{h}^{(1)}[s_1(g)]$ to get the multiplication of the element $h$ with word $w_1(g)$ to obtain $x_1$. Next access $\mathcal{A}_{x_1}^{(2)}[s_2(g)]$ to obtain $x_2=x_1w_2(g)$. Now repeat this process until we get the final result. The runtime is $\mathcal{O}(m)$ as we need to access the word arrays and the multiplication arrays $\mathcal{O}(m)$ times. 

\paragraph{Space Complexity:} The space use by the word arrays $\mathcal{W}$ is $\mathcal{O}(nm)$. The space used by the multiplication arrays $\mathcal{A}_g^{(i)}, ~ i \in [m], ~g\in G$ is $\mathcal{O}(2^l m n)$ as each array has length $2^l$. The overall space is $\mathcal{O}(nm+2^l  m n)$ which is $\mathcal{O}(2^l m n)$.\\

Recall, that $m=\lceil \frac{k}{l} \rceil$. From the above theorem we can assume that $m=c\log n$ for some constant $c$. If we set $l = \delta \log n$, then space used by our representation will be $\mathcal{O}(\frac{n^{1+\delta}}{\delta})$ words and the query time will be $\mathcal{O}(\frac{1}{\delta})$. Notice that as $l\geq 1$, we need $\delta\geq 1/{\log n}$.   
\end{proof}

\begin{corollary}
There is a representation of groups  such that multiplication query can be answered in $\mathcal{O}(\frac{\log n}{\log \log n})$ time using $\mathcal{O}(\frac{n(\log n)^2}{\log \log n})$ space .
\end{corollary}

\begin{proof}
Set $\delta = \mathcal{O}(\frac{\log \log n}{\log n})$ in Theorem~\ref{main theorem}. 
\end{proof}

\begin{corollary}
There is a representation of groups  such that multiplication query can be answered in $\mathcal{O}(\log n)$ time using $\mathcal{O}(n \log n)$ space.

\end{corollary}

\begin{proof}
Set $\delta = \mathcal{O}(\frac{1}{\log n})$ in Theorem~\ref{main theorem}. 
\end{proof}

\begin{corollary}

There is a representation of groups  such that multiplication query can be answered in $\mathcal{O}(1)$ time using $\mathcal{O}(n^{1.05})$ space.

\end{corollary}

\begin{proof}
Set $\delta = \frac{1}{20}$ in Theorem~\ref{main theorem}.
\end{proof}

\subsection{Deterministic Polynomial Time Preprocessing}\label{Deterministic Polynomial Time Preprocessing}
In this subsection we show that constructing the data structures (i.e., the word arrays and the multiplication arrays) takes polynomial time. It is easy to see that once we have a cube generating sequence for the input  group all the steps in constructing the data structures are polynomial time. The next theorem states that computing a cube generating sequence is also polynomial time. 

\begin{theorem}[\cite{babai1982representation}]
There is an algorithm that takes the Cayley table of a group $G$ as input and  computes a cube generating sequence  $(g_1,\ldots,g_k)$ of length $k=\mathcal{O}(\log n)$ in time $\mathcal{O}(n^2\log n)$ where $n=\vert G\vert$. 
\end{theorem}

\begin{proof}
The algorithm picks the elements $g_1,g_2,\ldots, g_k$ in stages. It starts by picking any arbitrary element $g_1$. Let $A_1=\{g_1\}$. Suppose at the end of stage $i-1$ the elements $g_1,g_2,\ldots, g_{i-1}$ have been picked. Let $A_{i-1}=\{g_1^{\epsilon_1}g_2^{\epsilon_2}\ldots g_{i-1}^{\epsilon_{i-1}}\mid \epsilon_1,\ldots,\epsilon_{i-1} \in \{0,1\}\}$. Next we describe how the algorithm picks $g_i$.

Let $X$ be a directed edge labeled graph with vertex set $G$. We put an edge from vertex $a$ to $b$ with label $g$ if $ag=b$. We consider the labels of the edges in the cut $E(A_{i-1}, G\setminus A_{i-1})=\{(a,b)\in E(X)\mid a\in A_{i-1} \textrm{ and } b\in G\setminus A_{i-1}\}$. 
Let $E_g(A_{i-1}, G\setminus A_{i-1})=\{(a,b)\in E(X)\mid a\in A_{i-1},  b\in G\setminus A_{i-1}, \textrm{ and } ag=b\}$. The algorithm picks an element $g$ such that $\vert E_g(A_{i-1}, G\setminus A_{i-1})\vert$ is maximized. Notice that we can compute $\vert E_g(A_{i-1}, G\setminus A_{i-1})\vert$ in time $\mathcal{O}(n)$. Thus, $g$ could be found in $\mathcal{O}(n^2)$ time. The algorithm sets $g_i=g$ and $A_i=\{g_1^{\epsilon_1}g_2^{\epsilon_2}\ldots g_{i}^{\epsilon_{i}}\mid \epsilon_1,\ldots,\epsilon_{i} \in \{0,1\}\}$.

~
 
\noindent
\emph{Claim :} Let $a_i=\vert A_i\vert$ for all $i$. Then $(n-a_i)\leq (n-a_{i-1})^2/ n$ where $n=\vert G\vert$. 

~

Assuming this claim, we observe that the fraction  $p_i=(n-a_i)/n$ of elements outside $A_i$ satisfies the inequality $p_i\leq p_{i-1}^2$. This gives us $p_{i}\leq p_1^{2^{i-1}}=(1-1/n)^{2^{i-1}}$. The algorithm stops when $p_k<1/n$. In fact, when $p_k<1/n$ there will be no element outside $A_k$ and $p_k$ will actually be $0$. In other words the algorithm stops for the largest value of $k$ such that $p_{k-1}\geq 1/n$. Now we know that $e^{-x}\geq 1-x$ for all real number $x$. Setting $x=1/n$ we obtain $e^{-1/n}\geq (1-1/n)$. Hence $e^{-2^{k-2}/n}\geq (1-1/n)^{2^{k-2}}\geq p_{k-1}\geq 1/n$. Hence, the largest value of $k$ with  $2^{k-2}/n\leq \ln n$ gives the number of stages of the algorithm. It is now easy to check that $k=\mathcal{O}(\log n)$. Therefore, the runtime of the algorithm is $\mathcal{O}(n^2\log n)$.  

~

\noindent
\emph{Proof of the Claim :} We note that $\vert E(A_{i-1},G\setminus A_{i-1})\vert = a_{i-1}(n-a_{i-1})$. This is because for each pair $(a,b)$ with $a\in A_{i-1}$ and $b\in G\setminus A_{i-1}$ there is exactly one edge from $a$ to $b$ and this edge is labeled with the element $a^{-1}b$. Thus, 
\[\vert E(A_{i-1},G\setminus A_{i-1})\vert = \sum_{g\in G}\vert E_g(A_{i-1},G\setminus A_{i-1})\vert=a_{i-1}(n-a_{i-1}).\] 

Hence, there is an element $g\in G$ such that $\vert E_g(A_{i-1},G \setminus  A_{i-1})\vert\geq a_{i-1}(n-a_{i-1})/n$. We further note that the edges in $E_g(A_{i-1},G\setminus A_{i-1})$ are vertex disjoint. Thus, $\vert A_{i}\vert \geq \vert A_{i-1}\vert +a_{i-1}(n-a_{i-1})/ n$. Hence, $a_{i} \geq  a_{i-1} +a_{i-1}(n-a_{i-1})/ n$. The claim follows from this. 
\end{proof}

%##########################################################################################
\section{Space Efficient Representations for Special Group Classes}
%##########################################################################################
In many of the results in this paper, group elements are treated as tuples. For example, if $\{g_1,\ldots,g_k\}$ is a cube generating set for a group $G$, $(\epsilon_1,\ldots,\epsilon_k)$ is a representation of the group element $g_1^{\epsilon_1}\cdots g_k^{\epsilon_k}$. For many of the data structures we design, we want a way of encoding these tuples which can be stored efficiently. We also want to retrieve the  group element from its encoding efficiently.

\textbf{Forward and Backward Map}: Let $G$ be a group, $c_1,\ldots,c_k$ be $k$ integers each greater than 1 with  $\prod_i c_i=\mathcal{O}(n)$ and $F:G\longrightarrow [c_1]\times\cdots\times[c_k]$ be an injective map. 
Suppose $F(g)=(\alpha_1,\ldots,\alpha_k)$. Let $\overline{b}_i$ be the $\ceil{\log c_i}$-bit binary encoding of $\alpha_i$ (possibly some $0$'s padded on the left to make it a $\ceil{\log c_i}$-bit string). 
The concatenation $\mathbf{b}=\overline{b}_1\ldots\overline{b}_k$ of the $\overline{b}_i$'s encodes $(\alpha_1,\ldots, \alpha_k)$. The encoding $\mathbf{b}$ can be stored in constant number of words as $\sum_i\ceil{\log c_i}=\mathcal{O}(\log n)$. Thus, $F$ can be stored in an array $\mathcal{F}$, indexed by the group elements using $\mathcal{O}(n)$ words by setting $\mathcal{F}[g]=\mathbf{b}$.  
We call this \emph{the forward map}. 
We also store an array $\mathcal{B}$, called \emph{the backward map}, of dimension $c_1\times\cdots \times c_k$ such that $\mathcal{B}[\alpha_1]\cdots[\alpha_k]=g$ if $F(g)=(\alpha_1,\cdots,\alpha_k)$.\footnote{If $(\alpha_1,\cdots,\alpha_k)\notin \text{Image}(F)$, then the value could be arbitrary.}
Finally we also store each $c_i$ in  separate words, which could be used to extract $(\alpha_1,\cdots,\alpha_k)$ from $\mathcal{F}[g]$ in $\mathcal{O}(\log n)$ time. Notice, that while the access to $\mathcal{F}$ is constant time, the access time for $\mathcal{B}$ is $\mathcal{O}(k)$ which is $\mathcal{O}(\log n)$.

\begin{lemma}\label{cyc-group}
There is a representation of cyclic groups such that multiplication operation can be performed using $\mathcal{O}(n)$ space in $\mathcal{O}(1)$ time.
\end{lemma}

\begin{proof}
 Let $G = \langle g \rangle$. Every element $g'$ in $G$ can be written as a $g' = g^i$ for some $ 0 \le i < |G|$. We use a  forward-map array $\mathcal{F}$ indexed by the elements of $G$ such $\mathcal{F}[g'] = i$ whenever $g' = g^i$ with $i \in \{0,\ldots,n-1\}$. Let $\mathcal{B}$ be  a backward-map array indexed by numbers from  $0$ to $n-1$, $\mathcal{B}[i] = g^i$.  Let $g_1$ and $g_2$ be the two elements to be multiplied. Let $g_1 = g^i$ and $g_2 = g^j$ such that $ 0\le i,j < n$ which can be obtained by accessing the forward-map array $\mathcal{F}$ at index $g_1$ and $g_2$ respectively. The result of the multiplication of $g_1$ and $g_2$ is $g_1g_2$ which is $g^ig^j =g^{(i+j)\%n}$. The computation $(i+j)\% n$ can be performed easily by first integer addition operation followed by a modulo operation. Let $k = (i+j)\%n$. Finally accessing the backward-map array $\mathcal{B}$ at index $k$ gives the final result. The data structure used here are  the forward-map array $\mathcal{F}$ and  the backward-map array $\mathcal{B}$. Both the arrays take  $\mathcal{O}(n)$ space and thus the overall space required is $\mathcal{O}(n)$. In query phase, we need one access the  forward-map array twice  followed by addition operation modulo operation and at the end one access the backward-map array once. 

\end{proof}

Many groups arise naturally as a semidirect product of its subgroups. In the following part we proved that if group $G$ is a semidirect product of its subgroups and if these subgroups admit  space efficient representation then $G$ also admits a space efficient representation.

Let $\mathcal{A}$ and $\mathcal{B}$ be two group classes.  
Let $ \mathcal{G}_{\mathcal{A},\mathcal{B}} = \{G \mid G =  A \rtimes_\varphi B, A\in \mathcal{A}, B\in \mathcal{B}, \text{and } \varphi \text{ is a homomorphism from } B \text{ to Aut}(A)\}$.

\begin{theorem}
Let $\mathcal{A}, \mathcal{B}$ be two group classes. Suppose we are given data structures $D_\mathcal{A}$ and $D_\mathcal{B}$ for group classes $\mathcal{A}$ and $\mathcal{B}$ respectively. Let $S(D_\mathcal{A}, m_1)$, $S(D_\mathcal{B}, m_2)$ denote the space required by the data structures $D_\mathcal{A}, D_\mathcal{B}$ to represent groups of order $m_1,m_2$ from $\mathcal{A}$, $\mathcal{B}$ respectively. Let $Q(D_\mathcal{A}, m_1), Q(D_\mathcal{B}, m_2)$ denote the time required by the data structures $D_\mathcal{A}, D_\mathcal{B}$ to answer multiplication queries for groups of order $m_1,m_2$ from $\mathcal{A},\mathcal{B}$ respectively.
Then there is a representation of groups in $\mathcal{G}_{\mathcal{A},\mathcal{B}}$ such that multiplication query can be answered in $\mathcal{O}(\log n + Q(D_\mathcal{A}, \vert A \vert) + Q(D_\mathcal{B}, \vert B \vert))$ time and $\mathcal{O}(n + S(D_\mathcal{A}, \vert A \vert) + S(D_\mathcal{B}, \vert B \vert))$ space.
\end{theorem}

\begin{proof}
First we describe the preprocessing phase. Given group $G$, finding  two groups $A,B$ and a homomorphism $\varphi$ such that $G = A \rtimes_\varphi B$ can be done in finite time where $\varphi : B \longrightarrow \text{Aut}(A)$. Without loss of generality, one can assume that elements of the group $A$ are numbered from $1$ to $\vert A \vert$. For each element $b \in B$, we store its image $\varphi(b) \in \text{Aut}(A)$ in the array $\mathcal{T}_b$ indexed by elements of group $A$. Let $\mathcal{T} = \{\mathcal{T}_b \mid b \in B\}$ be the set of $\vert B \vert$ arrays.

Now we move on to the querying phase. Let $g_1$ and $g_2$ be the two elements to be multiplied. Let $g_1 = (a_1,b_1)$ and $g_2 = (a_2,b_2)$ such that $a_1,a_2 \in A$ and $b_1,b_2 \in B$ which can be obtained using the forward map array. The result of the multiplication query $g_1g_2$ is $(a_1(\varphi(b_1)(a_2)),\,b_1b_2)$. The only \emph{non-trivial} computation involved here is computing $\varphi(b_1)(a_2)$, which can be obtained using array $\mathcal{T}_{b_1}$. Let $\mathcal{T}_{b_1}(a_2) = a_3$, then the result of the multiplication query $g_1g_2$ is $(a_1a_3,b_1b_2)$ the components of which can be computed using data structures for $A$ and $B$ respectively to obtain $(a_4, b_4)$ where $a_4 = a_1 a_3$ and $b_4 = b_1 b_2$. Finally using the backward map we can obtain the resultant element.

The data structures we use are -- forward-map array, $\vert B \vert$ many arrays $\mathcal{T}$ (each of size $\vert A \vert$), data structures for $B$ and $A$ and the backward map. Thus the overall space required is $\mathcal{O}(n + \vert B \vert \vert A \vert + S(D_\mathcal{A},\vert A \vert) + S(D_\mathcal{B},\vert B \vert) + n)$ which is $\mathcal{O}(n + S(D_\mathcal{A},\vert A \vert) + S(D_\mathcal{B},\vert B \vert))$.

The query time constitutes of the time required to get a representation of $g_1$ and $g_2$ as $(a_1, b_1)$ and $(a_2, b_2)$ respectively using the forward array, time required to compute $(a_1(\varphi(b_1)(a_2)),\,b_1b_2)$ using $\mathcal{T}_{b_1}$, time required to compute $a_1a_3$ and $b_1b_2$ and time required to access the backward-map array to obtain the resultant element. Thus the overall time required is $\mathcal{O}(1 + 1 + Q(D_{\mathcal{A}},\vert A \vert) + Q(D_{\mathcal{B}},\vert B \vert) + \log n)$ which is $\mathcal{O}(\log n + Q(D_{\mathcal{A}},\vert A \vert) + Q(D_{\mathcal{B}},\vert B \vert) )$. 
\end{proof}

Recall that $Z$-groups are groups which can be written as a semidirect product of two cycles. We now present a theorem which directly follows from the above theorem and Lemma~\ref{cyc-group}.

{
\renewcommand{\thetheorem}{\ref{Z-groups}}
\begin{theorem}
There is a representation of Z-groups such that multiplication operation can be performed using $\mathcal{O}(n)$ space in $\mathcal{O}(1)$ time.
\end{theorem}

}

  %##########################################################################
   %                                                    SIMPLE GROUPS

%##########################################################################
Simple groups serve as the building blocks for classifying finite groups. We next present a space efficient representation of simple groups.

{
\renewcommand{\thetheorem}{\ref{simple-groups}}

\begin{theorem}
There is a representation of simple groups such that multiplication operation can be performed using $\mathcal{O}(n)$ space in $\mathcal{O}(\log n)$ time.
\end{theorem}

}

\begin{proof}
The proof is divided into two cases. First assume that given simple group $G$ is abelian. It is easy to note that $G \cong\mathbb{Z}_p $, where $p$ is some prime, by using the fact that any abelian simple group is isomorphic to $\mathbb{Z}_{p}$, where $p$ is some prime. There is a representation of cyclic groups which answer a multiplication query in $\mathcal{O}(1)$ time using $\mathcal{O}(n)$ space (see Lemma~\ref{cyc-group}). We now assume that group $G$ is nonabelian. Babai, Kantor and Lubotsky \cite{BabaiKL89} proved that there is a constant $c$ such that every nonabelian finite simple group has a set $S$ of size at most 14 generators such that the diameter of the Cayley graph of $G$ with respect to $S$ is at most $c\log n$. Such a generating set can be found by iterating over all possible subsets of size 14. Let $G =\langle \mathcal{S} \rangle=\langle s_1,\cdots,s_{14}\rangle$. Each $g\in G$ can be represented by the edge labels in one of fixed shortest paths from the identity to $g$ in the Cayley graph. By the result of Babai, Kantor  and Lubotsky \cite{BabaiKL89} the length of the path is $\mathcal{O}(\log n)$. The edge labels are from $\{1,2,\ldots, 14\}$ indicating the generators associated with the edge. 
This representation of each element by the sequence of edge labels can be stored using a forward map. We also store a multiplication table $M$ of dimension $\vert G \vert\times [14]$. We set $M[g][i]=g g_i$. To multiply two elements $g$ and $h$ we consult the forward map for the representation of $h$ and then use $M$ to compute the $gh$ in $\mathcal{O}(\log n)$ time.

\end{proof}

%####################################################
                                         %   FARZON MODEL
                                            
                  %                         #################################

\section{Representation in the Model of Farzan and Munro}

In this section we use the model of computation defined by Farzan and Munro \cite{farzan2006succinct} for the succinct  representation of abelian groups. We describe the model briefly here.  For further details about this model, refer to \cite{farzan2006succinct}. Farzan and Munro use a Random Access Model (RAM) where a word is large enough to hold the value of the order of the input group $n$. The model also assumes the availability of bit-reversal as one of the native operations which can be performed in $\mathcal{O}(1)$ time.  

Given a group $G$, \emph{the compression algorithm} is defined as the process that takes $G$ as an input and outputs a \emph{compressed form} of $G$. 
 
The \emph{labeling} of elements of group $G$ (based on the compression) is a representation of the elements. Let  $A$ be an abelian group and  $t$ be the number of cyclic factors in the structural decomposition of $A$.  We denote by $\mathcal{L}_{A}: A \longrightarrow \mathbb{N}^t$ the labeling of elements as per Farzan and Munro's labeling \cite{farzan2006succinct}. 

We denote by \emph{outside user}, the entity responsible for the preprocessing operations such as compression, labeling etc. We denote by \emph{query processing unit}, the entity responsible for performing the actual multiplication. The query processing unit is responsible for storing the compressed form of the group $G$. The outside user is responsible for supplying to the query processing unit the labels of the group elements to be multiplied. The query processing unit returns the label of the result of the multiplication query. The space and time required in the compression and labeling phase is not counted towards the algorithm's space complexity and query time. 
Thus, in the following sections, we only consider the space and time consumed by the query processing unit.

\begin{theorem}[\cite{farzan2006succinct}]\label{farzan}
There is a representation of finite abelian group of order $n$ that uses constant number of words and answers multiplication queries in constant time.
\end{theorem}
\setcounter{theorem}{0}
Answering the question posed by \cite{farzan2006succinct}, we design data structures similar to the ones used in Theorem \ref{farzan}, for Hamiltonian groups and Z-groups. We also come up with a representation for groups which can be expressed as a semidirect product of an abelian group with a cyclic group.

\subsection{Hamiltonian groups }\label{hamiltonian}

{
\renewcommand{\thetheorem}{\ref{hamiltonian-munro}}

\begin{theorem}
There is a representation of Hamiltonian groups such that multiplication operation can be performed using $\mathcal{O}(1)$ space in $\mathcal{O}(1)$ time.
\end{theorem}
\addtocounter{theorem}{-1}
}

\begin{proof}

Let $G$ be a Hamiltonian group.  We know that $G$ can be decomposed as $G=Q_8 \times C$, where $Q_8$ is a quaternion group and $C$ is abelian (see Section~\ref{preliminarary}). The compressed form of group $G$ is same as the compressed form of the abelian group $C$.
Every element $g \in G$ has a representation of the form $(q,d)$ where $q \in Q_8$ and $d \in C$. The elements $q\in Q_8$ are assigned labels from the set $\{1,\dots,8\}$. Since $C$ is abelian, we use $\mathcal{L}_C(d)$ as the label for element $d \in C$. 
Since the order of $Q_8$ is constant, storing its entire Cayley table in some arbitrary but fixed representation requires $\mathcal{O}(1)$ space.  

Given two elements $r,s$ of $G$ such that $r=(q_1,d_1)$ and $s=(q_2,d_2)$ where $q_1,q_2 \in Q_8$ and $d_1,d_2 \in C$. The result of $rs$ is $(q_1q_2,d_1d_2)$. The multiplication of $q_1$ and $q_2$ can be computed in $\mathcal{O}(1)$ time using the stored Cayley table. After obtaining the labels $\mathcal{L}_C(d_1)$ and $\mathcal{L}_C(d_2)$ of elements $d_1$ and $d_2$ respectively, we can perform Farzan's multiplication algorithm to obtain the result of the multiplication of $d_1$ and $d_2$ in $\mathcal{O}(1)$ time. The overall space required is $\mathcal{O}(1)$ words. 
\end{proof}

\subsection{Z-groups}

We now consider Z-groups which are semidirect product of two cyclic groups. This group class contains the groups studied by Le Gall \cite{Gal}. We exploit the fact that every automorphism of a cyclic group is a cyclic permutation.

{
\renewcommand{\thetheorem}{\ref{Z-groups-munro}}

\begin{theorem}
There is a representation of Z-groups such that multiplication operation can be performed using $\mathcal{O}(1)$ space in $\mathcal{O}(1)$ time.
\end{theorem}
\addtocounter{theorem}{-1}
}

\begin{proof}

Let $G = C_m \rtimes_{\varphi} C_d$ be a Z-group where $\varphi : C_d \longrightarrow \text{Aut}(C_m)$ is a homomorphism and $C_m = \langle g \rangle$ and  $C_d = \langle h \rangle$. Without loss of generality, we assume that the elements of $C_m$ are numbered from the set $[m]$ in the natural cyclic order starting from $g$. Let $ \sigma_{j}:=\varphi(h^j)$. Let $\sigma_{j}(c)$ denote the image of the element $c \in C_m$ under the automorphism $\sigma_j$. 
In the compression process we first obtain a decomposition of $G$ as $C_m \rtimes_\varphi C_d$. The compressed form of group $G$ comprises of the two integers $m$ and $d$ and the compressed form of $\varphi$ which is $\sigma_1(g)$.
 
In the labeling phase, we label each element $t \in G$, such that $t = (g^i, h^j)$ as $(i, (\sigma_j(g), j))$ where $i \in [m]$ and $j \in [d]$.
Note that, with this labeling (computed by the outside user), representing any element from $G$ takes $\mathcal{O}(1)$ words.

In the querying phase, given two elements $r,s \in G$ such that $r=(g^{i_1},h^{j_1})$ and $s=(g^{i_2},h^{j_2})$ where $i_1,i_2 \in [m]$ and $j_1,j_2 \in [d]$. The result of $rs$ which is $(g^{i_1} \varphi(h^{j_1})(g^{i_2}),~ h^{j_1} h^{j_2}) = (g^{i_1}\sigma_{j_1}(g^{i_2}),~h^{j_1 + j_2})$. To compute $\sigma_{j_1}(g^{i_2})$, first obtain $\sigma_{j_1}(g)$ from the label of $r=(g^{i_1}, h^{j_1})$. Now to compute  $\sigma_{j_1}(g^{i_2})$ we need to perform one integer multiplication operation $(\sigma_{j_1}(g) \times {i_2}) \mathbin{\%} m$. Now the problem of multiplication reduces component-wise to the cyclic case. The multiplication query can thus be answered in $\mathcal{O}(1)$ time using $\mathcal{O}(1)$ space to store the orders of $C_m$ and $C_d$. 
\end{proof}

\subsection{Semidirect Product Classes }\label{semi-direct}

A natural way to construct nonabelian groups is by the extension of abelian groups. The groups which can be formed by semidirect product extension of abelian groups by cyclic groups has been studied by Le Gall \cite{Gal}. We denote $\mathcal{G}$ to be the class of groups which can be written as $G= A \rtimes C_m$, where $A$ is an abelian group and $C_m$ is a cyclic group. It is easy to see that the group class $\mathcal{G}$ is \emph{categorically larger} than abelian groups as it contains all abelian groups as well as some nonabelian groups.
Without loss of generality, assume that the elements of the group $A$ are numbered from $1$ to $ \vert A \vert$.

\begin{fact}
Any permutation can be decomposed as a composition of disjoint cycles \cite{dummit2004abstract}.
\end{fact}

\begin{lemma}\label{lem:fact}
Given an abelian group $A$, a permutation $\pi$ on the set $\{1,\dots,\vert A \vert \}$ and an element $g \in A$, there exists a representation of $\pi$ such that $\pi^{d}(g)$ for any element $g \in A$ and $d \in [m]$ can be computed in $\mathcal{O}(1)$ time using $\mathcal{O}(\vert A \vert)$ words of space.
\end{lemma} 

\begin{proof}

Let $\pi = \pi_1\circ \pi_2 \circ \cdots \circ \pi_l$ be the decomposition of $\pi$ into disjoint cycles $\pi_i, i \in [l]$. Such a decomposition of the input permutation $\pi$ can be computed in polynomial time. Let $\mathcal{C}_1,\ldots \mathcal{C}_l$ be arrays corresponding to the cycles $\pi_1,\ldots,\pi_l$ respectively. For any cycle $\pi_i$ store the elements of the cycle in $\mathcal{C}_i$ in the same order as they appear in $\pi_i$, starting with the least element of $\pi_i$. 
Construct an array $\mathcal{B}$ indexed by the elements of group $A$, storing for each $g \in A$, $\mathcal{B}[g] = (j,r),$ where $j \in [l]$ and $r \in \{ 0, \dots, \vert \mathcal{C}_j \vert -1 \}$ such that $\mathcal{C}_j[r] = g$. 
Now, in order to compute $\pi^{d}(g)$, first we obtain $j$ and $r$ from $\mathcal{B}$ such that the $g$ appears in the cycle $\pi_j$ at the $r$th index. Then we compute $r' := (r+d) \mathbin{\%} \vert \mathcal{C}_j \vert$ and finally return $\mathcal{C}_j[r']$. This requires $\mathcal{O}(1)$ time as the involved operations are one word operations. Note that we require overall $\mathcal{O}(\vert A \vert)$ space to store the arrays $\mathcal{C}_1, \ldots, \mathcal{C}_l$ as $\sum_i \vert\mathcal{C}_i \vert= \vert A \vert$, and the space required for $\mathcal{B}$ is also $\mathcal{O}(\vert A \vert)$. 
\end{proof}

{
\renewcommand{\thetheorem}{\ref{semi-munro}}

\begin{theorem}
There is a representation of groups $G \in \mathcal{G}$ such that multiplication query can be answered in $\mathcal{O}(1)$ time using $\mathcal{O}(\vert A \vert)$ space.
\end{theorem}
\addtocounter{theorem}{-1}
}

\begin{proof}

Let $G  \in \mathcal{G}$ be such that $G= A \rtimes_{\varphi} C_m$ where $\varphi : C_m \longrightarrow \text{Aut}(A)$ is a homomorphism. In the compression process, we first obtain the decomposition of group $G$ as $A \rtimes_\varphi C_m$. The compressed form of group $G$ comprises of the compressed form of group $A$, the integer $m$ and the space efficient  representation of the homomorphism $\varphi$ (described below). 

Let $C_m = \langle g \rangle$ and $ \pi :=\varphi(g)$. Note that $\pi$ is a bijection on the set $\{1,\ldots,\vert A \vert\}$. Using Lemma \ref{lem:fact} we can store $\pi$ in $\mathcal{O}(\vert A \vert)$ words, such that $\pi^d(a)$ for $a \in A$  can be computed in $\mathcal{O}(1)$ time. This forms the space efficient  representation of $\varphi$. Since the data structure used above is a part of the compressed form of group $G$, the query processing unit is responsible for storing it. We label each element $h \in G$ such that $h =(a,g^i)$ as $((\mathcal{L}_A(a),a), i)$ where $a \in A$, $c \in C_m$. 

This labeling requires $\mathcal{O}(1)$ words of space for each element $h \in G$. 
In the querying phase, given two elements $r = (a_1,c_1)$ and  $s=(a_2,c_2)$ of $G$ such that $a_1,a_2 \in A$ and $c_1,c_2 \in C_m$, the result of $rs$  is $(a_1(\varphi(c_1)(a_2)),c_1c_2)$. Let $c_1 = g^{k_1}$ and $c_2 = g^{k_2}$. Now $\varphi(c_1)(a_2)$ which is $\pi^{k_1}(a_2)$ can be computed using Lemma \ref{lem:fact} in $\mathcal{O}(1)$ time. Let $a_3 := \pi^{k_1}(a_2)$, then $rs = (a_1a_3, c_1c_2)$. Since the query processing unit is storing the labels of all the elements of $A$, it can obtain the label for the element $a_3$. After obtaining the labels $\mathcal{L}_A(a_1)$ and $\mathcal{L}_A(a_3)$ of elements $a_1$ and $a_3$ respectively, we can perform Farzan's multiplication algorithm to obtain the result of the multiplication of $a_1$ and $a_3$ in $\mathcal{O}(1)$ time. Let $c_3 = g^{k_3}$ be result of multiplication of $c_1$ and $c_2$. Then $k_3 = (k_1 + k_2) \mathbin{\%} m$ can be computed in $\mathcal{O}(1)$ time.

In the query processing unit, we are storing the elements of the group $A$ along with their labels, which takes $\mathcal{O}(\vert A \vert)$ words of space. The query processing unit also needs $\mathcal{O}(\vert A \vert)$ space to store the data structures from Lemma~\ref{lem:fact}. Thus the total space required is $\mathcal{O}(\vert A \vert)$.  

\end{proof}

\bibliographystyle{splncs04}

\bibliography{blb}

\begin{thebibliography}{10}
\providecommand{\url}[1]{\texttt{#1}}
\providecommand{\urlprefix}{URL }
\providecommand{\doi}[1]{https://doi.org/#1}

\bibitem{babai2016graph}
Babai, L.: Graph isomorphism in quasipolynomial time. In: Proceedings of the
  forty-eighth annual ACM symposium on Theory of Computing. pp. 684--697. ACM
  (2016). \doi{10.1145/2897518.2897542}

\bibitem{babai1982representation}
Babai, L., Erd{\"o}s, P.: Representation of group elements as short products.
  In: North-Holland Mathematics Studies, vol.~60, pp. 27--30. Elsevier (1982)

\bibitem{BabaiKL89}
Babai, L., Kantor, W.M., Lubotsky, A.: Small-diameter cayley graphs for finite
  simple groups. Eur. J. Comb. pp. 507--522 (1989).
  \doi{10.1016/S0195-6698(89)80067-8}

\bibitem{barrington1989bounded}
Barrington, D.A.: Bounded-width polynomial-size branching programs recognize
  exactly those languages in {NC$^1$}. Journal of Computer and System Sciences
  pp. 150--164 (1989). \doi{10.1016/0022-0000(89)90037-8}

\bibitem{Hal}
Carmichael, R.D.: Introduction to the Theory of Groups of Finite Order. GINN
  and Company (1937)

\bibitem{che}
Chen, L., Fu, B.: Linear and sublinear time algorithms for the basis of abelian
  groups. Theoretical Computer Science pp. 4110--4122 (2011).
  \doi{10.1016/j.tcs.2010.06.011}

\bibitem{dummit2004abstract}
Dummit, D.S., Foote, R.M.: Abstract algebra, vol.~3. Wiley Hoboken (2004)

\bibitem{erdos1965probabilistic}
Erd{\"o}s, P., R{\'e}nyi, A.: Probabilistic methods in group theory. Journal
  d'Analyse Math{\'e}matique pp. 127--138 (1965). \doi{10.1007/BF02806383}

\bibitem{eugene1993permutation}
Eugene, M.: Permutation groups and polynomial-time computation. In: Groups and
  Computation: Workshop on Groups and Computation, October 7-10, 1991. vol.~11,
  p.~139. American Mathematical Soc. (1993)

\bibitem{farzan2006succinct}
Farzan, A., Munro, J.I.: Succinct representation of finite abelian groups. In:
  Symbolic and Algebraic Computation, International Symposium, {ISSAC} 2006,
  Genoa, Italy, July 9-12, 2006, Proceedings. pp. 87--92. ACM (2006).
  \doi{10.1145/1145768.1145788}

\bibitem{Kav}
Kavitha, T.: Linear time algorithms for abelian group isomorphism and related
  problems. J. Comput. Syst. Sci. pp. 986--996 (Sep 2007).
  \doi{10.1016/j.jcss.2007.03.013}

\bibitem{Kay}
Kayal, N., Nezhmetdinov, T.: Factoring groups efficiently. In: International
  Colloquium on Automata, Languages, and Programming. pp. 585--596. Springer
  (2009). \doi{10.1007/978-3-642-02927-1\textunderscore49}

\bibitem{kumar1998property}
Kumar, S.R., Rubinfeld, R.: Property testing of abelian group operations (1998)

\bibitem{leedham1990collection}
Leedham-Green, C.R., Soicher, L.H.: Collection from the left and other
  strategies. Journal of Symbolic Computation pp. 665--675 (1990).
  \doi{10.1016/S0747-7171(08)80081-8}

\bibitem{Gal}
{Le\,Gall}, F.: {An Efficient Quantum Algorithm for Some Instances of the Group
  Isomorphism Problem}. In: 27th International Symposium on Theoretical Aspects
  of Computer Science. vol.~5, pp. 549--560 (2010).
  \doi{10.4230/LIPIcs.STACS.2010.2484}

\bibitem{luks1982isomorphism}
Luks, E.M.: Isomorphism of graphs of bounded valence can be tested in
  polynomial time. Journal of computer and system sciences pp. 42--65 (1982).
  \doi{10.1016/0022-0000(82)90009-5}

\bibitem{magnus2004combinatorial}
Magnus, W., Karrass, A., Solitar, D.: Combinatorial group theory: Presentations
  of groups in terms of generators and relations. Courier Corporation (2004)

\bibitem{Sar}
Qiao, Y., {Sarma M.N.}, J., Tang, B.: {On Isomorphism Testing of Groups with
  Normal Hall Subgroups}. In: 28th International Symposium on Theoretical
  Aspects of Computer Science (STACS 2011). pp. 567--578 (2011).
  \doi{10.4230/LIPIcs.STACS.2011.567}

\bibitem{seress2003permutation}
Seress, {\'A}.: Permutation group algorithms, vol.~152. Cambridge University
  Press (2003)

\bibitem{sims1994computation}
Sims, C.C.: Computation with finitely presented groups, vol.~48. Cambridge
  University Press (1994)

\bibitem{Vik}
Vikas, N.: An $\mathcal{O}(n)$ algorithm for abelian $p$-group isomorphism and
  an $\mathcal{O}(n \log n)$ algorithm for abelian group isomorphism. Journal
  of Computer and System Sciences pp.~1--9 (1996). \doi{10.1006/jcss.1996.0045}

\bibitem{Wil}
Wilson, J.B.: Existence, algorithms, and asymptotics of direct product
  decompositions, {I}. Groups Complexity Cryptology pp. 33--72 (2012).
  \doi{10.1515/gcc-2012-0007}

\end{thebibliography}

\appendix

\end{document}